\documentstyle[12pt,fleqn]{article}
\def\soc{{\rm C}_{60}}
\def\rug{{\rm C}_{70}}

\def\la{\langle}
\def\ra{\rangle}
\def\beeq{\begin{equation}}
\def\eneq{\end{equation}}
\def\beeqa{\begin{eqnarray}}
\def\eneqa{\end{eqnarray}}

\addtocounter{section}{-1}
\begin{document}

\begin{center}

\vspace{2in}

{\large {\bf{Metallic and insulating states\\
in two-dimensional C$_{\bf 60}$-polymers
} } }



\vspace{1cm}

{\rm Kikuo Harigaya\footnote[1]{E-mail address:
harigaya@etl.go.jp; URL: http://www.etl.go.jp/People/harigaya/.}
}\\

\vspace{1cm}

{\sl Fundamental Physics Section,\\
Electrotechnical Laboratory,\\
Umezono 1-1-4, Tsukuba, Ibaraki 305, Japan}

\vspace{1cm}

(Received~~~~~~~~~~~~~~~~~~~~~~~~~~~~~~~~~~~)
\end{center}

\Roman{table}

\vspace{1cm}

\noindent
{\bf Abstract}\\
Variations in the band structures of the two-dimensional
$\soc$-polymers are studied, when $\pi$-con\-ju\-ga\-tion
conditions are changed.  We investigate the rectangular
and triangular polymers, in order to discuss metal-insulator
transitions, using a semiempirical model with the
Su-Schrieffer-Heeger type electron-phonon interactions.
We find that electronic structures change among direct-gap
insulators and the metal, depending on the degree of
$\pi$-conjugations in the rectangular polymer.  The triangular
polymer changes from the indirect gap insulator to the metal
as the $\pi$-conjugations increase.  High pressure experiments
could observe such pressure-induced metal-insulator transitions.



\pagebreak

\section{Introduction}

Recently, it has been found that the linear (one-dimensional)
$\soc$-polymers are realized in alkali-metal doped $\soc$ crystals:
$A_1\soc$ ($A=$K, Rb) [1-4], and their solid state properties
are intensively investigated.  One electron per one $\soc$ is
doped in the polymer chain.  It seems that Fermi surfaces exist
in high temperatures, but the system shows antiferromagnetic
correlations in low temperatures [1].  The $\soc$-polymer has
lattice structures where $\soc$ molecules are arrayed in a linear
chain.  The bonds between $\soc$ are formed by the [2+2] cycloaddition
mechanism.  And, quite recently, the other new phases of $\soc$ systems
which are synthesized at high pressures have been reported [5,6].
The two-dimensional polymer structures [6,7] have been proposed
and a tight-binding calculation [8] has been already reported.
It has been discussed [8] that their lattice structures can be
explained by the calculation.  The two kinds of lattice structures
-- rectangular and triangular structures -- are present in the
two-dimensional polymers.  The structures are shown in Fig. 1.
Figure 1(a) is for the rectangular polymer, and Fig. 1(b) displays
the triangular polymer.  There are four membered rings between
neighboring $\soc$ molecules.  These rings are the results of the
[2+2] cycloaddition.

In this paper, we shall study variations in the band structures
of the two-dimensional $\soc$-polymers when the $\pi$-conjugation
conditions are changed.  We assume that the bonding states
between $\soc$ might be changed easily possibly by applying high
pressures.  As illustrated in Fig. 3 of the ref. [2], there
are three candidates for the classical bonding structures around
the four membered ring.  The first case (A) is that there are only
weak van der Waals interactions between $\soc$, and the classical
double bonds remain as in the isolated $\soc$ molecule.  The second
case (B) is that all the bonds of the four membered ring are the
single bond, so they are $\sigma$-like in the quantum chemistry.
All the four-membered rings in Fig. 1 are formed by the $\sigma$-bonds
in this case.  Then, the third case (C) is that the bonds which connect
the neighboring molecules are the double bonds, and the bonds
derived from the isolated $\soc$ -- the bonds, $\langle 1,2 \rangle$,
$\langle 3,4 \rangle$, $\langle 5,6 \rangle$, $\langle 7,8 \rangle$,
of Fig. 1(a) -- are destroyed completely.  Here, $\langle i,j \rangle$
indicates the pair of the neighboring $i$ and $j$th atoms.
In this case, the bonds between $\soc$ -- the
bonds, $\langle 1,5 \rangle$, $\langle 2,6 \rangle$,
$\langle 3,8 \rangle$, $\langle 4,7 \rangle$, of Fig. 1(a) -- have the
$\sigma$- as well as $\pi$-characters.  In other words, the degree of
the $\pi$-conjugations between the molecules becomes maximum.  The
main purpose of this paper is to propose a model which can deal with
changes of $\pi$-conjugation conditions among the above three cases,
and to look at electronic band structures of the two types of
$\soc$-polymers.  We note that the operator at the lattice
sites of the four membered rings is one of the relevant linear
combinations of the effective $\sigma$-like components, assuming
a possibility of local $\sigma$-conjugations at the four
membered rings.  The similar assumption of the $\sigma$-conjugation
has been used in Si-based polymers, for example, in ref. [9].
We, however, use the term ``$\pi$-conjugation" for simplicity
in this paper, because the local $\sigma$-conjugations can
be regarded as a part of the global $\pi$-conjugations which
are extended over the system.

We propose a semiempirical tight-binding model analogous to
the Su-Schrieffer-Heeger (SSH) model [10] of conjugated polymers.
The electronic structures may depend sensitively upon the
$\pi$-conjugation conditions even in the neutral polymer,
because the several bonds connecting neighboring molecules are
largely distorted, and the mixing between $\sigma$- and $\pi$-orbitals
will change only by slight change of the bond structures [2].  We would
like to study effects of the change of the $\pi$-conjugation
conditions by introducing a phenomenological parameter in a tight-binding
model.  The model is an extension of the SSH-type model which
has been applied to $\soc$ [11,12] and $\rug$ [12,13] molecules.
The model is solved with the assumption of the adiabatic approximation,
and band structures are reported in order to discuss metal-insulator
changes by varying $\pi$-conjugations.

We will conclude that the electronic structures change among
direct-gap (indirect-gap) insulators and the metal, depending on
the degree of $\pi$-conjugations.  In the rectangular polymer, the
electronic structures change from the direct gap insulator
with the gap at the $\Gamma$ point, through the metal, to the
insulator with the direct gap at the K point, as increasing
the $\pi$-conjugation between $\soc$ molecules.  The high
pressure experiments may be able to change $\pi$-conjugation
conditions between $\soc$ molecules, and the electronic structure
changes could be observed. It may seem that the transition to metals
in high pressure is a general fact which is common to various
materials [14].  However, in the present theory, the rectangular
polymer changes from an insulator through a metal, and finally
changes into an insulator again.  This reentrant behavior is
specific to the present $\soc$-polymer systems, and is not a
guess from the general knowledge about high pressure effects.
In the triangular polymer, there is an indirect energy gap
at the intermediate $\pi$-conjugations.  And, the system changes
into a metal at the maximum $\pi$-conjugation.

In the next section, our tight-binding model is introduced and
an idea of changing $\pi$-conjugation conditions is discussed.
The sections 3 is devoted to numerical results.
The paper is closed with a summary in section 4.

\section{Model}

We would like to apply an SSH-type model to the two-dimensional
$\soc$-polymers.  In the previous works [11-13], we have proposed
the extended SSH model to $\soc$ and $\rug$ molecules.  In $\soc$,
all the carbon atoms are equivalent, so it is a good approximation
to neglect the mixing between $\pi$- and $\sigma$-orbitals.  The
presence of the bond alternation and the energy level structures
of the neutral $\soc$ molecule can be quantitatively described by
the calculations within the adiabatic approximation.  In $\rug$,
the molecular structure becomes longer, meaning that the degrees of
the mixing between $\pi$- and $\sigma$-characters are different
depending on carbon sites.  In this respect, the extended SSH model
does not take account of the differences of the mixings.  However,
it has been found [12,13] that qualitative characters of the
electronic level structures are reasonably calculated when the
extended SSH model is applied to the $\rug$.  This is a valid
approach because the $\sigma$-orbitals can be simulated by the
classical harmonic springs in the first approximation.

In this paper, we assume the same idea that the lattice
structures and the related molecular orbitals of
each $\soc$ molecule in the $\soc$-polymers can be described
by the SSH-type model with the hopping interactions for the
$\pi$-orbitals and the classical springs for the $\sigma$-orbitals.
However, the mixings between the $\pi$- and $\sigma$-orbitals
near the eight bonds, $\langle i,j \rangle$ ($i,j=1 - 8$),
shown in Fig. 1(a) are largely different from those of regions
far from the four bonds.  We shall shed light on this special
character of bondings between the neighboring $\soc$.
Electronic structures would be largely affected by changes
of $\pi$-conjugation conditions (or local $\sigma$-conjugations
as in Si-based polymers [9]) around the four bonds.
We shall introduce a semiempirical parameter $a$ as shown
in the following hamiltonian:
\beeqa
H_{\rm pol} &=&  a \sum_{l,\sigma}
{\sum_{\langle i,j \rangle}}^{'}
(- t + \alpha y_{l,\langle i,j \rangle} )
( c_{l,i,\sigma}^\dagger c_{l+1,j,\sigma} + {\rm h.c.} ) \\ \nonumber
&+&  (1-a) \sum_{l,\sigma}
{\sum_{\langle i,j \rangle}}^{''}
(- t + \alpha y_{l,\langle i,j \rangle} )
( c_{l,i,\sigma}^\dagger c_{l,j,\sigma} + {\rm h.c.} ) \\ \nonumber
&+& \sum_{l,\sigma} \sum_{\langle i,j \rangle = {\rm others}}
(- t + \alpha y_{l,\langle i,j \rangle} )
( c_{l,i,\sigma}^\dagger c_{l,j,\sigma} + {\rm h.c.} ) \\ \nonumber
&+& \frac{K}{2} \sum_i \sum_{\langle i,j \rangle} y_{l,\langle i,j \rangle}^2,
\eneqa
where $t$ is the hopping integral of the system without the
bond alternations in the isolated $\soc$ molecule; $\alpha$ is the
electron-phonon coupling constant which changes the hopping
integral linearly with respect to the bond variable
$y_{l,\langle i,j \rangle}$, where $l$ means the $l$th
molecule and $\langle i,j \rangle$ indicates the pair of
the neighboring $i$ and $j$th atoms; the atoms with $i=1 - 8$
of the rectangular polymer are shown by numbers in Fig. 1(a), and
the atoms with $i=1 - 12$ of the triangular polymer are in Fig. 1(b);
the other $i$ and $j$ in the third column of eq. (1) label
the nonnumbered atoms in the same molecule; $c_{l,i,\sigma}$
is an annihilation operator of the $\pi$-electron
at the $i$th site of the $l$th molecule with spin $\sigma$;
the sum is taken over the pairs of neighboring atoms;
and the last term with the spring constant $K$
is the harmonic energy of the classical spring simulating the
$\sigma$-bond effects.  Note that the sum with the prime is
performed over $\langle i,j \rangle = \la 1,5 \ra, \la 2,6 \ra,
\la 3,8 \ra, \la 4,7 \ra$ for the rectangular polymer, and
it is performed over $\la i,j \ra = \la 1,8 \ra, \la 2,7 \ra,
\la 3, 10 \ra, \la 4,9 \ra, \la 5, 12 \ra, \la 6, 11 \ra$
for the triangular polymer.  The sum with the double prime
is performed over $\la i,j \ra = \la 1,2 \ra, \la 3,4 \ra,
\la 5,6 \ra, \la 7,8 \ra$ for the rectangular polymer, and
it is performed over $\la i,j \ra = \la 1,2 \ra, \la 3,4 \ra,
\la 5,6 \ra, \la 7,8 \ra, \la 9,10 \ra, \la 11,12 \ra$
for the triangular polymer.

As stated before, the parameter $a$ controls the strength of
$\pi$-conjugations between neighboring molecules.  When $a=1$,
the $\sigma$-bondings between atoms, 1 and 2, 3 and 4, 5 and 6,
7 and 8, in Fig. 1(a) are completely broken and the orbitals
would become like $\pi$-orbitals.  The bond between the atoms 1 and 5
and the equivalent other bonds become double bonds.  This is
the case (C).  As $a$ becomes smaller, the $\pi$-conjugation
between the neighboring molecules decreases, and the $\soc$
molecules become mutually independent.  In other words, the
interactions between molecules become smaller in the intermediate $a$
region.  In this case (B), the operator $c_{l,i,\sigma}$ at the lattice
sites of the four membered rings is one of the relevant linear
combinations of the effective $\sigma$-like components.  Here,
we assume a possibility of local $\sigma$-conjugations at the four
membered rings.  The similar assumption of the $\sigma$-conjugation
has been used in Si-based polymers, for example, in ref. [9].
When $a=0$ [the case (A)], the $\soc$ molecules are completely
isolated each other.  The band structures of the $\soc$-polymers
will change largely depending on the $\pi$-conjugation conditions.
We discuss this effects in the next section.

The present unit cell consists of one $\soc$ molecule for
the two polymers shown in Fig. 1.   Using the lattice periodicity,
we suppress the index $l$ of the bond variable $y_{l,\langle i,j \rangle}$.
In other words, all the molecules in the polymers are assumed
to have the same lattice structures.  The bond variables are
determined by using the adiabatic approximation in the real space.
The same numerical iteration method as in ref. [12] is used here.
We will change the parameter, $a$, within $0.5 \leq a \leq 1.0$.
We consider the neutral case so the electron number is 60
for one $\soc$.  The other parameters, $t=2.1$eV,
$\alpha = 6.0$eV/\AA, and $K = 52.5$eV/\AA$^2$, give
the energy gap 1.904eV and the difference between the short and
bond lengths 0.04557\AA\ for an isolated $\soc$ molecule.
We shall use the same parameter set here.

\section{Numerical results}

We show band structures of the rectangular polymers for the
$\pi$-conjugation conditions, $a = 0.5$, 0.7, 0.8, 0.9, and
1.0, as the representative cases in Fig. 2.  In Figs. 2(a) and (b),
the highest occupied states and the lowest unoccupied states
are named as ``HOMO" and ``LUMO".  At $a=0.5$,
there is a direct energy gap at the $\Gamma$ point.  The system
is an insulator.  As discussed in ref. [8], the density of
states of the system in the room pressures is
large and the band widths are narrow.  The parameter
$a=0.5$ seems to be large for the rectangular polymer in room
temperatures and pressures [6].  However, we shall report the
relatively large $a$ cases, because our central interests are
the behaviors of the system in the large conjugations which could
be realized in high pressures.  As the parameter $a$ increases,
the HOMO moves upward and the LUMO goes downward.  The energy
gap decreases as shown for $a=0.7$ in Fig. 2(b).  The deep and
high energy bands do not move so apparently because they have
large amplitudes of wavefunctions in the central parts of
the $\soc$ balls.  In Fig. 2(c) ($a=0.8$), the crossing of the
HOMO and LUMO of the smaller $a$ cases occurs.  There is Fermi
surfaces at about -0.3eV, and the system is metallic.
The crossing becomes more apparent for $a=0.9$ as shown
in Fig. 2(d).  At the maximum conjugations with $a = 1.0$,
the direct energy gap appears again at the K point, as shown
in Fig. 2(e).  The system changes into an insulator again.

Why do such the reentrant changes take place?  In order to
discuss the reason, we show the magnitudes of wavefunctions
at the $\Gamma$ point for (a) $a=0.5$ and (b) $a=1.0$,
in Fig. 3.  The nonequivalent sites with
respect to symmetries are labelled as A-I,
as shown in Fig. 1(a).  The wavefunctions can be taken
as real, so we use this convention.  The HOMO is shown by
white bars, and the LUMO is displayed by the black bars.
In Fig. 3(a), the HOMO has the negligible amplitude
at the site E, and the LUMO is near zero at the site I.
In contrast, as in Fig. 3(b), the LUMO is negligible
at the site E and the HOMO is near zero at the site I.
This fact indicates that the symmetries of the HOMO
and LUMO are reversed in the two insulators with the
small and large $a$'s.  Therefore, the crossing of the
HOMO and LUMO should occur at the intermediate $a$
in the present model.  This is the origin of the metallic
band structures.  Applying high pressures might decrease the
distances between $\soc$ molecules and thus increase $a$.
The insulating system changes into a metal and
then an insulator again.  This behavior is specific to the
rectangular $\soc$-polymers, and it seems a quite interesting
finding.

Next, we shall discuss band structures of the triangular
$\soc$-polymer shown in Fig. 1(b).  There are six nonequivalent
sites, and they are named as A-F.  As the qualitative features
of the band structure changes are similar to those of the rectangular
polymer, we show band structures only for (a) $a=0.5$ and
(b) $a=1.0$, in Fig. 4.  For $a=0.5$, the HOMO has the maximum
at the $\Gamma$ point, and the LUMO has the minimum at the
K point.  The system is an indirect gap insulator.
For $a = 1.0$, there are Fermi surfaces at about -0.1eV.
Therefore, a metallic transition could be observed
in the triangular polymer, too.  In this system, a reentrant
behavior is not found in the large $a$ region.  This might come
from the geometry difference between the rectangular
and triangular polymers.

\section{Summary}

We have studied the variations of the band structures in
rectangular and triangular phases of the two-dimensional
$\soc$-polymers.  We have changed the con\-ju\-ga\-tion
conditions between molecules.  A semiempirical model with
SSH-type electron-phonon interactions has been proposed.
The band structures have been shown, in order to discuss
metal-insulator changes.  A possibility of observing
electronic structure changes in high pressure experiments,
which may increase $\pi$-conjugations between $\soc$ molecules,
has been pointed out.  The reentrant behavior in the rectangular
polymer, which is special to the $\soc$ systems, is a quite
interesting finding.  We have discussed that the metallic
state is the result of the crossing of the bands, relating
with the symmetry properties of wavefunctions.

\pagebreak
\begin{flushleft}
{\bf References}
\end{flushleft}

\noindent
$[1]$ O. Chauvet, G. Oszl\`{a}nyi, L. Forr\'{o}, P. W. Stephens,
M. Tegze, G. Faigel and A. J\`{a}nossy,
Phys. Rev. Lett. 72 (1994) 2721.\\
$[2]$ P. W. Stephens, G. Bortel, G. Faigel, M. Tegze,
A. J\`{a}nossy, S. Pekker, G. Oszlanyi and L. Forr\'{o},
Nature 370 (1994) 636.\\
$[3]$ S. Pekker, L. Forr\'{o}, L. Mihaly and A. J\`{a}nossy,
Solid State Commun. 90 (1994) 349.\\
$[4]$ S. Pekker, A. J\`{a}nossy, L. Mihaly, O. Chauvet,
M. Carrard and L. Forr\'{o}, Science 265 (1994) 1077.\\
$[5]$ Y. Iwasa, T. Arima, R. M. Fleming, T. Siegrist, O. Zhou,
R. C. Haddon, L. J. Rothberg, K. B. Lyons, H. L. Carter Jr.,
A. F. Hebard, R. Tycko, G. Dabbagh, J. J. Krajewski, G. A. Thomas
and T. Yagi, Science 264 (1994) 1570.\\
$[6]$ M. M\'{u}\~{n}ez-Regueiro, L. Marques, J. L. Hodeau,
O. B\'{e}thoux and M. Perroux, Phys. Rev. Lett. 74 (1995) 278.\\
$[7]$ G. Oszlanyi and L. Forr\'{o}, Solid State Commun. 93 (1995) 265.\\
$[8]$ C. H. Xu and G. E. Scuseria, Phys. Rev. Lett. 74 (1995) 274.\\
$[9]$ T. Hasegawa, Y. Iwasa, H. Sunamura, T. Koda, Y. Tokura,
H. Tachibana, M. Matsumoto and S. Abe, Phys. Rev. Lett.
69 (1992) 668.\\
$[10]$ W. P. Su, J. R. Schrieffer and A. J. Heeger, Phys. Rev. B
22 (1980) 2099.\\
$[11]$ K. Harigaya, J. Phys. Soc. Jpn. 60 (1991) 4001.\\
$[12]$ K. Harigaya, Phys. Rev. B 45 (1992) 13676.\\
$[13]$ K. Harigaya, Chem. Phys. Lett. 189 (1992) 79.\\
$[14]$ E. Wigner and H. B. Huntington, J. Chem. Phys. 3 (1935) 764.\\

\pagebreak

\begin{flushleft}
{\bf Figure Captions}
\end{flushleft}

\mbox{}

\noindent
Fig. 1.  The crystal structures of the two-dimensional
$\soc$-polymers.  The rectangular polymer is shown in (a),
and the triangular polymer is displayed in (b).
The labels, A-I, indicate carbon atoms which are not
equivalent due to the symmetries.  These labels are
used in Fig. 3.  The carbon sites which constitute the
four membered rings are named with numbers.
In (a), the sites 2 (and 6) overlap with the sites 1 (and 5),
respectively, and cannot be seen from the front.
The parenthesis mean the overlapped sites.

\mbox{}

\noindent
Fig. 2.  The band structures of the rectangular $\soc$-polymer
of the cases (a) $a = 0.5$, (b) 0.7, (c) 0.8, (d) 0.9, and
(e) 1.0.  In (a), (b), and (e), the highest fully occupied
band is named as ``HOMO", and the lowest empty band as ``LUMO".

\mbox{}

\noindent
Fig. 3.  The magnitudes of the wavefunctions of
the HOMO and the LUMO at the $\Gamma$ point
for (a) $a=0.5$ and (b) 1.0, in the rectangular
polymer.  The nonequivalent sites are labelled as A-I,
as shown in Fig. 1(a).  The HOMO is shown by the
white bars, and the LUMO is displayed by the black bars.

\mbox{}

\noindent
Fig. 4.  The band structures of the triangular
$\soc$-polymer of the cases (a) $a = 0.5$ and (b) 1.0.
In (a), the highest occupied band is named as ``HOMO",
and the lowest empty band as ``LUMO".

\end{document}